\documentclass[aps,prb,twocolumn,superscriptaddress,showpacs,floatfix,amsmath,amssymb]{revtex4-1}
\usepackage{graphicx}

\begin{document}

\title{Quantum Critical Fluctuations in the Heavy fermion compound Ce(Ni$_{0.935}$Pd$_{0.065}$)$_2$Ge$_2$}

\author{C. H. Wang}
\affiliation{Oak Ridge National Laboratory, Oak Ridge, Tennessee 37831, USA}
\affiliation{University of California, Irvine, California 92697, USA}

\author{L. Poudel}
\affiliation{Oak Ridge National Laboratory, Oak Ridge, Tennessee 37831, USA}
\affiliation{University of Tennessee, Knoxville, Tennessee 37996, USA}

\author{A. E. Taylor}
\affiliation{Oak Ridge National Laboratory, Oak Ridge, Tennessee 37831, USA}

\author{J. M. Lawrence}
\affiliation{University of California, Irvine, California 92697, USA}

\author{A. D. Christianson}
\affiliation{Oak Ridge National Laboratory, Oak Ridge, Tennessee 37831, USA}
\affiliation{University of Tennessee, Knoxville, Tennessee 37996, USA}

\author{S. Chang}
\affiliation{National Institute of Standards and Technology, Gaithersburg, Maryland 20899-6102, USA}

\author{J. A. Rodriguez-Rivera}
\affiliation{National Institute of Standards and Technology, Gaithersburg, Maryland 20899-6102, USA}
\affiliation{University of Maryland, College Park, Maryland 20742, USA}

\author{J. W. Lynn}
\affiliation{National Institute of Standards and Technology, Gaithersburg, Maryland 20899-6102, USA}

\author{A. A. Podlesnyak}
\affiliation{Oak Ridge National Laboratory, Oak Ridge, Tennessee 37831, USA}

\author{G. Ehlers}
\affiliation{Oak Ridge National Laboratory, Oak Ridge, Tennessee 37831, USA}

\author{R. E. Baumbach}
\affiliation{Los Alamos National Laboratory, Los Alamos, New Mexico 87545, USA}
\affiliation{National High Magnetic Field Laboratory, Tallahassee, FL 32310, USA}

\author{E. D. Bauer}
\affiliation{Los Alamos National Laboratory, Los Alamos, New Mexico 87545, USA}

\author{K. Gofryk}
\affiliation{Los Alamos National Laboratory, Los Alamos, New Mexico 87545, USA}
\affiliation{Idaho National Laboratory, Idaho Falls, ID 83415, USA}

\author{F. Ronning}
\affiliation{Los Alamos National Laboratory, Los Alamos, New Mexico 87545, USA}

\author{K. J. McClellan}
\affiliation{Los Alamos National Laboratory, Los Alamos, New Mexico 87545, USA}

\author{J. D. Thompson}
\affiliation{Los Alamos National Laboratory, Los Alamos, New Mexico 87545, USA}

\date{\today}

\begin{abstract}

Electric resistivity, specific heat, magnetic susceptibility, and inelastic neutron scattering experiments were performed on a single crystal of the heavy fermion compound
Ce(Ni$_{0.935}$Pd$_{0.065}$)$_2$Ge$_2$ in order to study the spin fluctuations near an antiferromagnetic (AF)
quantum critical point (QCP).  The resistivity and the specific heat coefficient for $T \leq$ 1 K exhibit
the power law behavior expected for a 3D itinerant AF QCP ($\rho(T) \sim T^{3/2}$ and
$\gamma(T) \sim \gamma_0 - b T^{1/2}$). However, for 2 $\leq T \leq$ 10 K, the susceptibility and specific heat vary as $log T$ and the resistivity varies linearly with temperature. Furthermore, despite the fact that the resistivity and specific heat  exhibit the non-Fermi liquid behavior expected at a QCP, the correlation length, correlation time, and staggered susceptibility of the spin fluctuations remain finite at low temperature. We suggest that these deviations from the divergent behavior expected for a QCP may result from alloy disorder.

\end{abstract}


\maketitle

The nature of the quantum phase transition, where a quantum critical point (QCP) separates a magnetic
state from a nonmagnetic Fermi liquid (FL) state at temperature $T$=0, is one of the most fundamental
questions for strongly correlated electron systems.  Heavy fermion compounds, where the QCP is accessed by tuning
between antiferromagntic (AF) and FL states through application of non-thermal control parameters, such as
pressure or chemical doping ($\delta$=$P, x$), provide several good examples for studying the quantum phase
transition\cite{Lohneysen}. When the Kondo temperature $T_K$ is finite at the QCP then the phase transition
involves formation of spin density waves (SDW). These arise in the renormalized Fermi liquid that results from correlated
hybridization of the $f$-electrons with the conduction electrons.

The critical fluctuations near an AF critical point are characterized by a correlation time $\tau$, a correlation
length $\xi$, and a staggered susceptibility $\chi(Q_N)$, all of which should diverge at the QCP. The
dynamic exponent $z$ relates the correlation time to the correlation length via $\tau \propto \xi^z$. In
the traditional spin fluctuation theory of such an itinerant AF (SDW-type) $T=$ 0 transition\cite{Millis, Hertz},
the behavior near the QCP depends on both the physical dimension $d$ and the dynamic exponent $z$.
When the effective dimension $d+z$ is greater than the upper critical dimension $d^* =$ 4,  the
spin fluctuations exhibit Gaussian behavior at the QCP. For an AF transition, such theory makes a central assumption that $z=$ 2, so that Gaussian behavior is expected at a three dimensional SDW-type QCP. When implemented via the renormalization
group (RG) approach\cite{Hertz,Millis} or the self-consistent renormalization approach (SCR)\cite{Moriya}, the theory predicts that the resistivity varies as $T^{3/2}$, the linear coefficient of specific heat as $C/T = \gamma_0 - bT^{1/2}$, and the susceptibility $\chi$ as $T^{-3/2}$. The inverse staggered susceptibility $\chi(Q_N)^{-1}$ and the inverse correlation time $\Gamma(Q_N) \sim 1/\tau$ time are expected to vary as $T^{3/2}$.
Such theory works reasonably well to describe the behavior near the QCP of both the bulk properties of many heavy
fermion systems\cite{Stewart} and inelastic neutron scattering in alloys of CeRu$_2$Si$_2$\cite{Kadowaki1,Knafo2}.

\begin{figure}[t]
\centering
\includegraphics[width=0.45\textwidth]{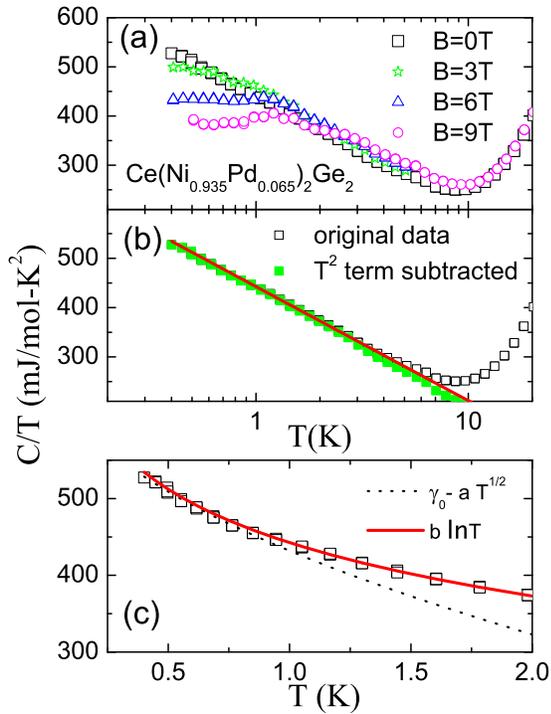}
\caption{\label{fig1} (color online) (a) The low temperature specific heat coefficient of Ce(Ni$_{0.935}$Pd$_{0.065}$)$_2$Ge$_2$ at several magnetic fields.  (b) Showing the effect of subtracting the $T^2$ term. (c) The specific heat coefficient at temperatures below 2 K. The solid red line in panels (b) and (c) represents $log T$ behavior; the dashed black line in panel (c) represents  $T^{1/2}$ behavior.}
\vspace*{-3.5mm}
\end{figure}

\begin{figure}[t]
\centering
\includegraphics[width=0.45\textwidth]{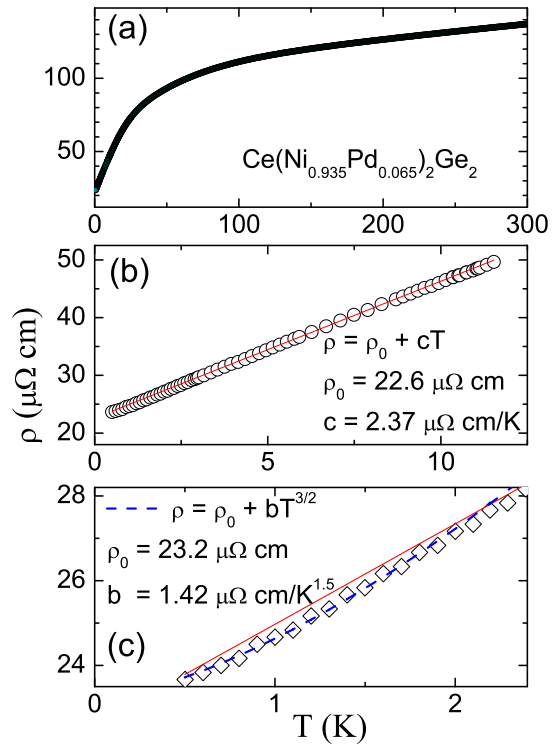}
\caption{\label{fig2} (color online) a) The resistivity of Ce(Ni$_{0.935}$Pd$_{0.065}$)$_2$Ge$_2$ over the temperature range 2 - 300 K. b) In the range 2-10 K the resistivity is linear with temperature (red line). c) Below 2 K, the resistivity varies as $\rho(T) \sim \rho(0) + b T^{3/2}$ (dashed line).}
\vspace*{-3.5mm}
\end{figure}

 At the lowest temperatures, the tetragonal compound CeNi$_2$Ge$_2$ exhibits heavy Fermi liquid behavior,
 with $C/T \simeq$ 380 mJ/mol K$^2$\cite{Koerner} and $T_K$ $\simeq$30 K\cite{Knopp}. Inelastic neutron
 spectra show two excitations\cite{Kadowaki3} centered in momentum space near \emph{Q} = (1/2 1/2 0)
 and (0 0 3/4). The energy of these excitations decreases with temperature down to 2 K; at lower
 temperatures, the characteristic energy remains constant at a value 0.75 meV. This compound can be
 driven to a QCP by alloying Pd onto the Ni site. According to the published phase diagram\cite{Knebel}, the critical concentration for the
 QCP in Ce(Ni$_{1-x}$Pd$_{x}$)$_2$Ge$_2$ is $x_c =$ 0.065. Compared to CeNi$_2$Ge$_2$, the specific
 heat coefficient of Ce(Ni$_{0.935}$Pd$_{0.065}$)$_2$Ge$_2$
is enhanced to the value $C/T \simeq$ 700 mJ/mol K$^2$ and the Kondo temperature is reduced to
$T_K \simeq$ 17 K\cite{Kuwai}. This finite Kondo temperature, plus the similarities of this compound to alloys of  CeRu$_2$Si$_2$ where SDW-type QCP behavior is observed\cite{Kadowaki1,Knafo2}, suggests that Ce(Ni$_{0.935}$Pd$_{0.065}$)$_2$Ge$_2$ should exhibit the critical behavior expected for an itinerant AF QCP.

In this work, we present results of resistivity, susceptibility, specific heat, and  inelastic neutron scattering measurements performed on a
single crystal of the alloy Ce(Ni$_{0.935}$Pd$_{0.065}$)$_2$Ge$_2$. The results show  that $z$ = 2,
and the resistivity for $T  \leq$ 2 K and specific heat for $T \leq$ 1 K behave as expected for a quantum phase transition in an itinerant antiferromagnet. However, the correlation length, correlation time, and staggered susceptibility do not diverge at low temperature. We argue that this fundamental discrepancy from theory reflects the important role of alloy disorder near the QCP.

A 5 gram single crystal was grown by the Czochralski method using $^{58}$Ni to avoid  the large incoherent
scattering of neutrons by natural Ni.  Characterization by energy-dispersive x-ray spectroscopy showed that the sample had a uniform alloy concentration. The specific heat (Fig. 1 a) was found to be identical to that which was presented in a
previous paper\cite{Kuwai} for $x=$ 0.06.

The neutron scattering experiments were performed on the Cold Neutron Chopper Spectrometer (CNCS)\cite{Ehlers} at the Spallation Neutron Source (SNS) at Oak Ridge National Laboratory. Additional measurements were carried out on the multi-axis crystal spectrometer (MACS)\cite{MACS} at the NIST Center for Neutron Research (NCNR). The crystal was aligned in the ($H H L$) scattering plane. On these spectrometers, a large (discretized) volume of momentum-energy space can be mapped by rotating the sample through a series of angles, and constant $Q$ or constant $E$ spectra can be obtained by interpolation. (We note that throughout the paper, the values of $Q$ are given in reciprocal lattice units (r.l.u.), which are defined via $a_0/2\pi Q_x$, $a_0/2\pi Q_y$, and $c_0/2\pi Q_z$ where $a_0$ and $c_0$ are the lattice constants.)

The results for the specific heat, resistivity, and susceptibility are shown in Figs. 1, 2, and 3. Below 1 K, the specific heat can indeed be fit as $\gamma(T) = \gamma (0) -a T^{1/2}$ and below 2 K the resistivity varies as $\rho(T) = \rho(0) + b T^{3/2}$. For 2 $\leq T \leq$ 10 K, however, the resistivity is linear with temperature and the variation of the susceptibility with $T$ is closer to logarithmic than to  $T^{-3/2}$ behavior. In addition, the specific heat coefficient can be fit to a logarithmic temperature dependence from 0.4 to 10 K.  All this behavior is quite characteristic of heavy fermion compounds where the QCP is attained by alloying.\cite{Lohneysen,Stewart}

\begin{figure}[t]
\centering
\includegraphics[width=0.5\textwidth]{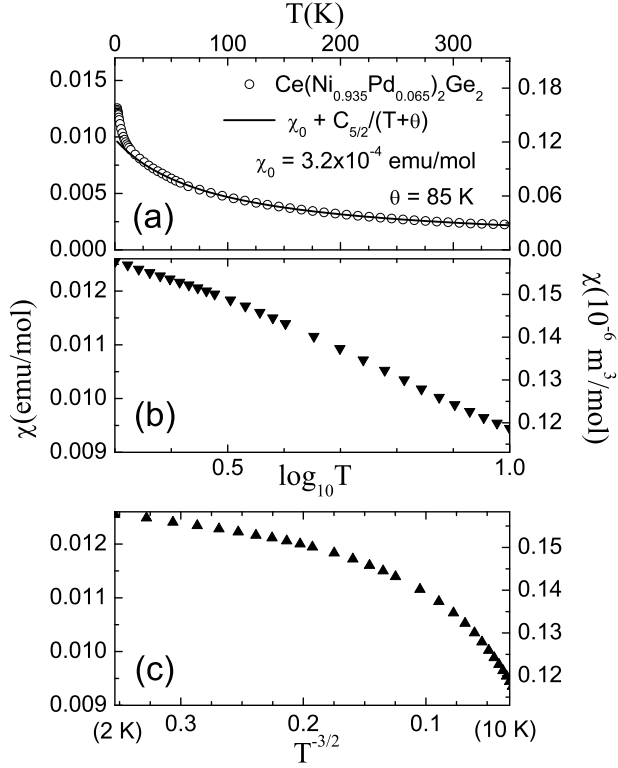}
\caption{\label{fig3} (color online) a) The susceptibility of Ce(Ni$_{0.935}$Pd$_{0.065}$)$_2$Ge$_2$ over the temperature range 2 - 350 K, along with the high temperature Curie-Weiss fit. ($C_{5/2}$ is the cerium free ion Curie constant.) b) and c) Between 2 and 10 K, the susceptibility is better fit by $log T$ than by $T^{-3/2}$ behavior.}
\vspace*{-3.5mm}
\end{figure}

\begin{figure}[t]
\centering
\includegraphics[width=0.5\textwidth]{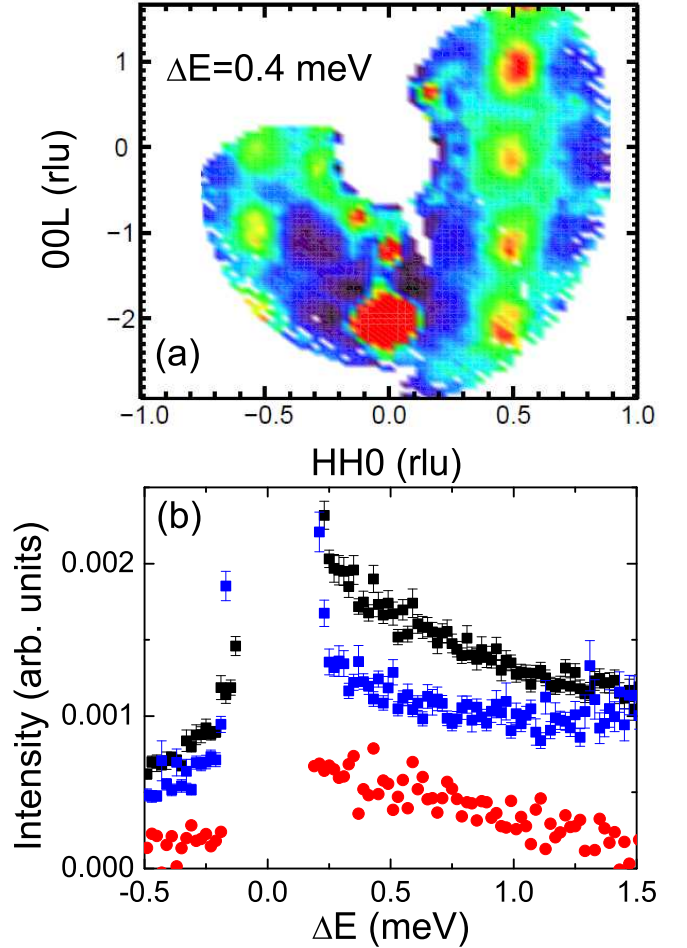}
\caption{\label{fig4} (color online) (a) Map of intensity in the ($H H L$) plane obtained on MACS
at $\Delta E$=0.4 meV and $T$ = 0.5 K.
(b) Constant-$Q$ scans at $Q_N=$ (1/2 1/2 0) (black symbols) and (3/4 3/4 0) (blue symbols) as determined on CNCS at 0.27 K. We subtract the latter as an estimate of the background to determine the magnetic scattering near $Q_N$ (red symbols). (Error bars throughout represent one standard deviation.)}
\vspace*{-3.5mm}
\end{figure}

\begin{figure}[t]
\centering
\includegraphics[width=1\columnwidth]{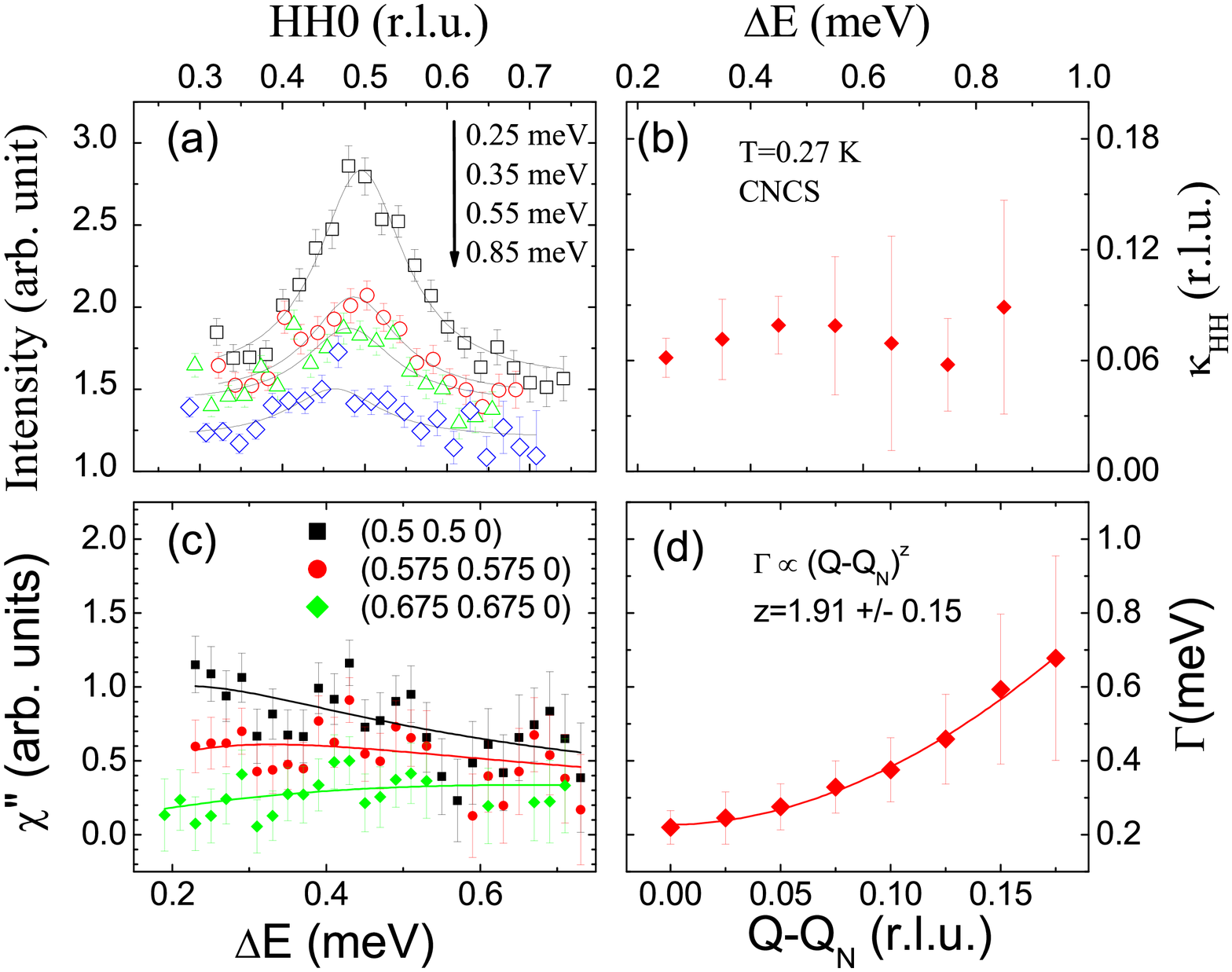}
\caption{\label{fig5} (color online) (a) Constant $E$ scans along the ($H$ $H$ 0) direction at several
energy transfers as measured on CNCS at 0.27 K. The solid lines are fits to Eq. 2. (b) The
dependence of the parameter $\kappa$ on energy transfer, determined from a). The measured coherence length is essentially independent
of energy transfer.(c) Intensity versus
energy transfer at several $Q$ in the vicinity of $Q_N$ at 0.27 K. The lines are fits
to Eq. 1. (d) The inverse correlation time $\Gamma$ versus $Q - Q_N$, as derived from (c). The solid
line is a fit to the relation $\Gamma = \Gamma_0 + (Q-Q_N)^z$.}
\vspace*{-3.5mm}
\end{figure}

For the neutron measurements, the energy transfer was fixed at
$\Delta E$= 0.4 meV to map the Brillouin zone at 0.5 K in order to determine the location in
$Q$-space of the AF fluctuations. Results measured at 0.27 K on CNCS are identical.
Fig. 4 a shows that the excitation is centered at the (1/2 1/2 0) position,
equal to the reported antiferromagnetic zone center $Q_N$ = (1/2 1/2 0.99) for 10\% Pd doping
where the alloy is magnetically ordered\cite{0.1Pd}.

Two typical
energy scans are presented in Fig. 4 b. The scans centered at the wavevector $Q_N$ = (1/2 1/2 0) show additional scattering compared to those at wavevectors such as (3/4 3/4 0), which lie outside the critical (bright) region in $Q$-space, i.e. in the blue region of Fig. 4 a.
We use the latter spectrum to estimate the background, which we subtract from the (1/2 1/2 0)
scattering to determine the dynamic susceptibility at $Q = Q_N$.

We  fit the resulting data
to the form \\\\
$\chi\prime\prime (\Delta E, Q) \sim \chi\prime (Q) \Gamma(Q) \dfrac{ \Delta E}{\Delta E^2 + \Gamma^2(Q)}$. (1)\\\\
Here, when $Q$ = $Q_N$, $\chi\prime$ is the staggered susceptibility $\chi(Q_N)$ and $\Gamma$ is related
to the critical correlation lifetime by $\tau \propto \Gamma^{-1}(Q_N)$. The inverse correlation
length $ \kappa^{-1} \propto \xi$ can be determined by examining the $Q$ dependence of $\chi\prime(Q)$:\\\\
$\chi\prime  (Q) \sim  \chi(Q_N) \dfrac{\kappa}{(Q-Q_N)^2 + \kappa^2}$. (2) \\\\
To obtain $\kappa$ from fits of the data to equations (1) and (2), we mapped intensity versus $Q$ near
$Q_N =$ (1/2 1/2 0) for several fixed energy transfers in the range of 0.25 meV to 1 meV (Fig. 5 a).
The width in $Q$ of the scattering does not change significantly for different $\Delta E$ in this range,
and fits of the ($H$ $H$ 0) or (1/2 1/2 $L$) cuts to equation (2)  show that the $\kappa$ values remain
the same within the experimental resolution (Fig. 5 b). Given this, $\kappa$ can be obtained at any
temperature by a fit of equation (2) to a $Q$-cut at a single energy, as in Fig. 6 b.

\begin{figure}[t]
\centering
\includegraphics[width=0.5\textwidth]{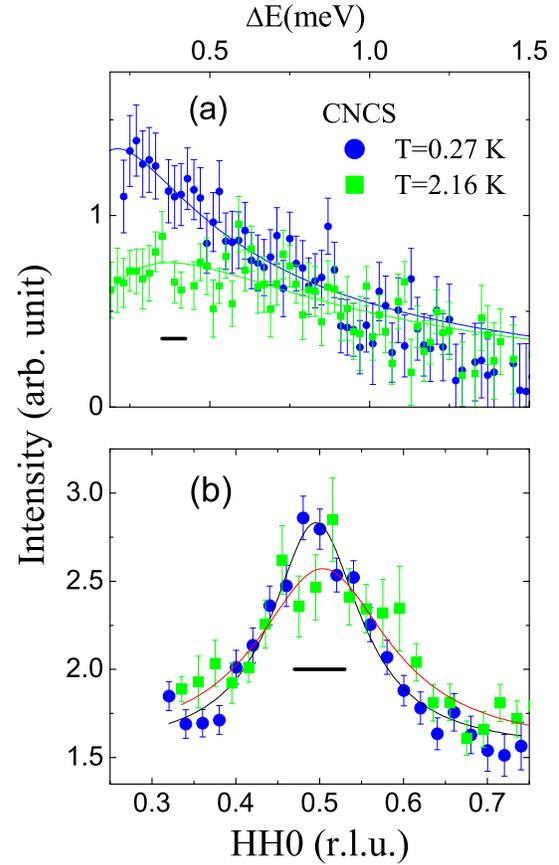}
\caption{\label{fig6} (color online) (a) Energy scans at $Q_N=$ (1/2 1/2 0) as measured on CNCS at 0.27 and 2.16 K. (b) Constant $E$ scans for $Q$ along the ($H H$ 0) direction for $\Delta E$ = 0.2 meV at $T=$ 0.27 and 2.16 K. The solid lines are fits to Eqs. 1 and 2. The horizontal lines in each panel represent the instrument resolution (FWHM). }
\vspace*{-3.5mm}
\end{figure}

\begin{figure}[t]
\centering
\includegraphics[width=0.5\textwidth]{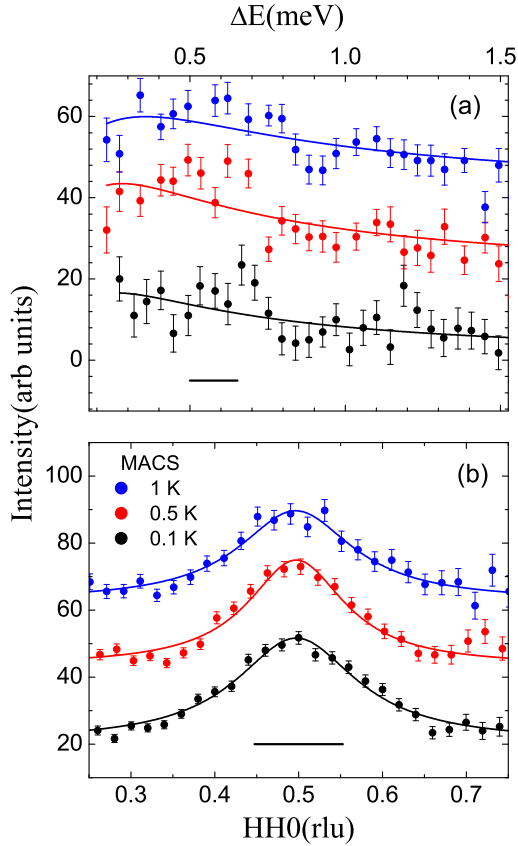}
\caption{\label{fig7} (color online) (a) Energy scans at $Q_N=$ (1/2 1/2 0) as measured on MACS at 0.1, 0.5, and 1 K. (b) Constant $E$ scans for $Q$ along the ($H H$ 0) direction for $\Delta E$ = 0.4 meV at $T=$ 0.1, 0.5 and 1 K. The solid lines are fits to Eqs. 1 and 2. The horizontal lines in each panel represent the instrument resolution (FWHM). The data at 0.5 K (1 K) have been shifted vertically by 20 (40) units in each panel.}
\vspace*{-3.5mm}
\end{figure}

\begin{figure}[t]
\centering
\includegraphics[width=1\columnwidth]{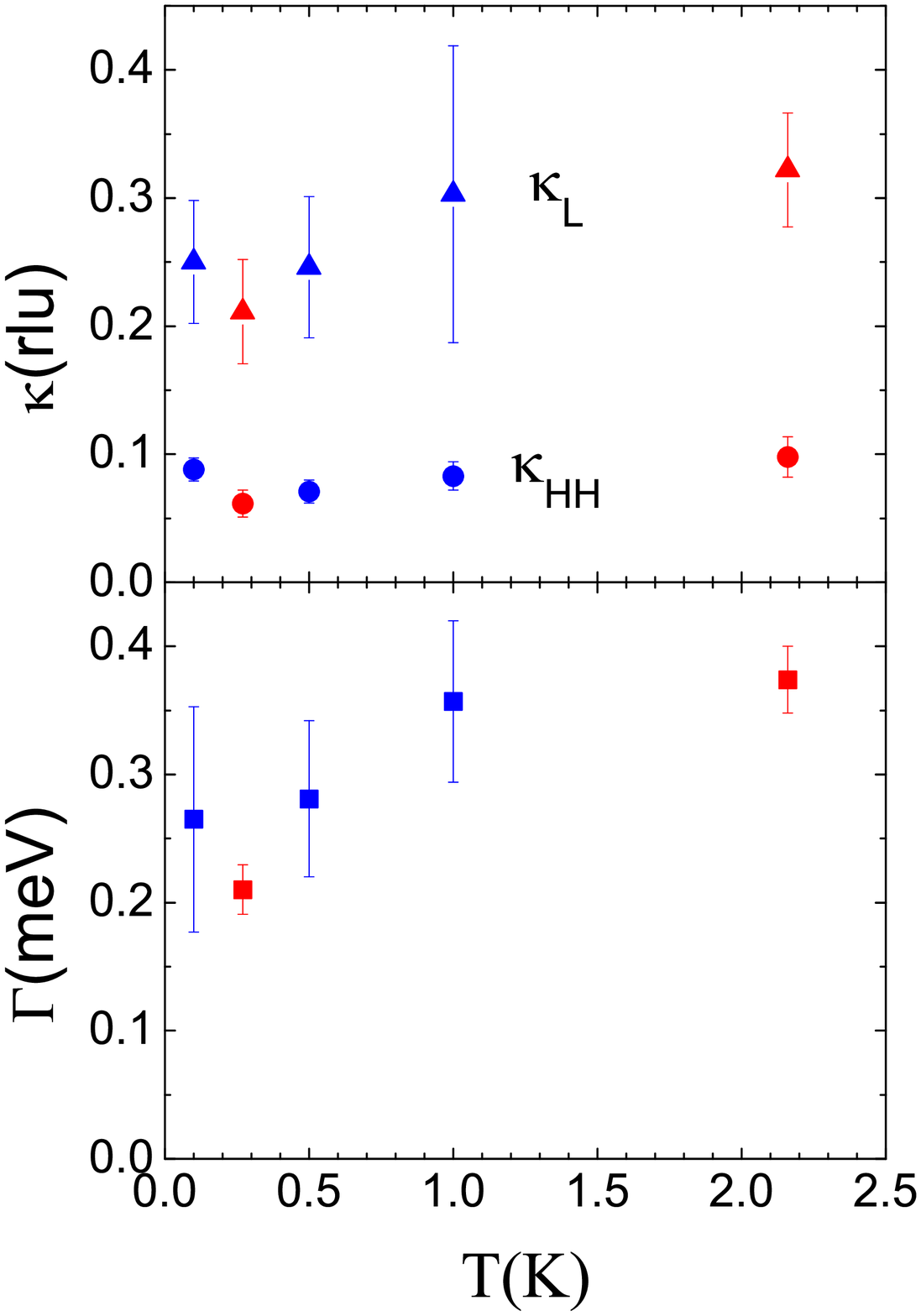}
\caption{\label{fig8} (color online) (a) Inverse correlation length $\kappa$ in the $HH$ and $L$ directions versus temperature.(b) The inverse correlation time $\Gamma$ versus temperature. Red symbols represent data from CNCS; blue symbols are for data from MACS.}
\vspace*{-3.5mm}
\end{figure}

The correlation lifetime $\tau$ is related to the correlation length $\xi$ via $\tau \propto \xi^z$, and the dynamic exponent $z$ is also expected to govern the dispersion relation $\Gamma \sim (Q - Q_N)^z$. In Figs. 5 c and d we examine the dispersion of the critical fluctuations
by fitting the spectra at several values of $Q$ in the vicinity of $Q_N$, determining $\Gamma$
and plotting $\Gamma$ vs $(Q - Q_N)$. The result  demonstrates that the dynamic exponent has a
value very close to 2.

The energy dependence of the spectra at $Q = Q_N$ and the $Q$ dependence for $\Delta E=$ 0.2 meV are plotted in Fig. 6 and 7. Fitting these data to Eqs. 1 and 2, we determine the temperature dependence of $\kappa$ and $\Gamma$ which we plot in Fig. 8. Both $\kappa$ and $\Gamma$ should vanish at $T=$ 0 at the QCP, whereas we observe finite values. The staggered susceptibility, which is proportional to the area under the spectra shown in Figs. 6 a and 7 a, also remains finite. Given the lattice constants $a=$ 4.15 {\AA} and  $c=$  9.85 {\AA},
the $T=$ 0 values $\kappa_{HH} =$ 0.06 r.l.u. and $\kappa_L =$ 0.21 r.l.u. correspond
to a correlation length of nearly 50 {\AA} in both the $H H$ and $L$ directions. Similarly, the value $\Gamma=$ 0.2 meV corresponds to a correlation time of 2 x 10$^{-11}$ seconds.

The divergences in the dynamic susceptibility that are expected at a QCP have not been observed experimentally in any heavy fermion alloy system. A finite correlation length at a QCP has been observed for Ce(Fe$_{1-x}$Ru$_x$)$_2$Ge$_2$\cite{Montfrooij}. Other Ce-122 alloys, such as Ce$_{1-x}$La$_x$Ru$_2$Si$_2$\cite{Knafo2}, CeRu$_{2-x}$Rh$_x$Si$_2$\cite{Kadowaki1},
and CeCu$_2$Si$_2$\cite{Arndt} also exhibit finite correlation times. The results show that $\Gamma$ varies as $\Gamma(0) + T^{3/2}$ in the critical region of these alloys. (It is highly probable that such behavior is valid in Ce(Ni$_{1-x}$Pd$_x$)$_2$Ge$_2$ at higher temperatures than reported here.) The finite values of  $\Gamma(0)$ that have been observed in the three above-mentioned systems are in the range 0.2-0.3 meV, similar to the value of 0.2-0.25 meV that we observe for Ce(Ni$_{0.935}$Pd$_{0.065}$)$_2$Ge$_2$.

The most obvious explanation for this lack of divergence of the spin
fluctuation lifetime is a deviation of the alloy concentration $x$ from the critical
value $x_c$, as proposed by Kadowaki \textit{et al.}\cite{Kadowaki1}.  On the other hand, if the actual alloy parameter $x$ for our sample were less than the critical value, then the specific heat coefficient should saturate to a constant value at low temperature, as seen for our sample at finite field (Fig. 1 a); if greater than $x_c$ then there should be an indication of an AF transition in the specific heat and susceptibility. However, our sample and indeed all the alloys mentioned above do not show such signatures of deviation from $x = x_c$ but show precisely the non-Fermi liquid behavior
for the specific heat and resistivity expected at the critical concentration. Given this, it is possible that the lack of divergences of the fluctuation lifetime, correlation length, and staggered susceptibility is not due to a deviation from the critical concentration, and needs to be taken seriously.

Other authors have noticed this lack of divergence and have offered alternative explanations. Knafo \textit{et al.}\cite{Knafo2} have proposed
that the finite $\chi (Q_N)$ and $\Gamma$ seen at the QCP in Ce$_{1-x}$La$_x$Ru$_2$Si$_2$ might arise because the
transition is weakly first order. To the best of our knowledge, however, there have been no observations of hysteresis or other indications of two-phase, first-order behavior in these compounds.

Alternatively, Montfrooij\cite{Montfrooij} has attempted to explain the finite correlation
length observed at the QCP in Ce(Fe$_{1-x}$Ru$_{x}$)$_2$Ge$_2$ in terms of alloy disorder. The underlying idea is that of Kondo disorder. The QCP occurs when the Kondo temperature becomes small enough compared to the RKKY exchange that magnetism can be stabilized. Due to the fact that the Kondo temperature is exponentially related to the 4f/conduction hybridization, statistical variations in the local cerium environment that affect the hybridization can lead to a significant distribution of Kondo temperatures near the QCP. The effect on the dynamic susceptibility will be that of an enhanced inhomogeneity rounding, which can explain the finite values of the correlation length and correlation time. It has also been shown\cite{Miranda,Graf,Booth} that such Kondo disorder can lead to the logarithmic temperature dependence of the specific heat coefficient and susceptibility and to a linear temperature dependence for the resistivity, as we observe in our sample for 2 $\leq T \leq$ 10 K.

In Montfrooij's view\cite{Montfrooij}, the emergence of ordered magnetism must be understood in the context of percolation theory. In the simplest version of this idea, there will be two cerium environments. Cerium atoms that are neighbors to Pd solute atoms will have stable local moments ($T_K =$ 0). The rest of the cerium atoms will have a finite value of Kodno temperature, and the local moment will disappear for $T < T_K$. The latter will create a background of SDW-type short range order, with finite lifetime and coherence length. Long range magnetic order will occur when the length of the networks of magnetic moments becomes macroscopic. At the QCP, these percolating networks will continue to co-exist with the regions of short range order. While the inelastic spectrum of the percolating network will exhibit critical divergences, the spectral weight will be dominated by the non-diverging regions of short range order. This would explain the saturation of the divergences in the dynamic susceptibility to finite values. We note that each Pd solute atom has four cerium near neighbors, so that it is plausible that the percolation limit can be reached for an alloy concentration as small as $x=$ 0.065.

A more sophisticated version of this theory considers the full set of possible cerium environments that can occur in a random alloy, and hence combines a distribution of $T_K$ (i.e. Kondo disorder) with percolation theory. This is the realm of Griffiths phase physics.\cite{CastroNeto} When the Harris criterion $\nu d<$ 2 is satisfied, alloy disorder is expected to cause the critical behavior to cross over at sufficiently low temperature from SDW-type behavior to that of the Griffiths phase. Since the mean field exponent $\nu =$ 1/2 for the correlation length is expected when $z+d>$ 4, the Harris criterion  is satisfied for 3D SDW-type systems. Hence it should apply for Ce(Ni$_{0.935}$Pd$_{0.065}$)$_2$Ge$_2$, for which the presence of alloy disorder is confirmed by the magnitude of the residual resistivity in Fig. 2.

On a low temperature scale, the SDW-type behavior should cross over to a critical behavior where the disorder plays a dominant role. Given the strict $E/T$ scaling expected in the Griffiths phase, the low temperature of the phase, the low moment expected near the QCP, and the low spectral weight of the spectrum of the percolating moments, this behavior should be extremely difficult to observe experimentally in the dynamic susceptibility.

In conclusion, we have shown experimentally that the dynamic exponent has the value $z$ = 2 near
the QCP in Ce(Ni$_{0.935}$Pd$_{0.065}$)$_2$Ge$_2$. This result as well as the behavior of the low temperature resistivity and specific heat are consistent with the expectations for a quantum transition in a 3D itinerant AF spin fluctuation
system. The low temperature correlation length, correlation time, and staggered susceptibility, however, remain finite at the lowest temperatures measured, suggesting the importance of alloy disorder. Future work is needed to determine the critical exponents more precisely and to discover whether there is a crossover to a
different critical behavior closer to the QCP.

We thank Chris Stock for his assistance in the measurement at NIST, Min-Nan Ou for assistance at SNS,
and Cristian Batista for helpful conversations. Research at ORNL was sponsored
by the Laboratory Directed Research and Development Program of Oak Ridge National Laboratory, and was
supported by the Scientific User Facilities Division Office of Basic Energy Sciences, DOE. Work at
UC Irvine and Los Alamos National Laboratory was supported by the U.S. Department of Energy,
Office of Basic Energy Sciences, Division of Materials Sciences and
Engineering; the work at UC-Irvine was funded under Award DE-FG02-03ER46036. This work utilized facilities supported
in part by the National Science Foundation under Agreement No. DMR-0944772.

\end{document}